\documentclass[aps,prl,twocolumn,showpacs,superscriptaddress,groupedaddress, showkeys]{revtex4}  

\usepackage{graphicx}  
\usepackage{dcolumn}   
\usepackage{bm}        
\usepackage{amssymb}   
\usepackage{color}
\usepackage{epstopdf}

\begin{document}

\title{Giant Rashba Splitting in CH$_3$NH$_3$PbBr$_3$ Organic-Inorganic Perovskite}

\author{Daniel~Niesner} 
\email{daniel.niesner@fau.de}
\author{Max Wilhelm}
\affiliation{Lehrstuhl f\"ur Festk\"orperphysik, Friedrich-Alexander-Universit\"at Erlangen-N\"urnberg (FAU), Staudtstr.~7, 91058~Erlangen, Germany}

\author{Ievgen Levchuk} 
\author{Andres Osvet}
\author{Shreetu Shrestha}
\author{Miroslaw Batentschuk}
\affiliation{Institute of Materials for Electronics and Energy Technology~(I-MEET), Department of Materials Science and Engineering, Friedrich-Alexander-Universit\"at Erlangen-N\"urnberg (FAU), Martensstrasse~7, 91058~Erlangen, Germany}

\author{Christoph Brabec}
\affiliation{Institute of Materials for Electronics and Energy Technology~(I-MEET), Department of Materials Science and Engineering, Friedrich-Alexander-Universit\"at Erlangen-N\"urnberg (FAU), Martensstrasse~7, 91058~Erlangen, Germany}
\affiliation{Bavarian Center for Applied Energy Research (ZAE Bayern), Haberstrasse 2a, 91058 Erlangen, Germany}

\author{Thomas Fauster}
\affiliation{Lehrstuhl f\"ur Festk\"orperphysik, Friedrich-Alexander-Universit\"at Erlangen-N\"urnberg (FAU), Staudtstr.~7, 91058~Erlangen, Germany}


\date{\today}

\begin{abstract}

As they combine decent mobilities with extremely long carrier lifetimes, organic-inorganic perovskites have opened a whole new field in opto\-electronics. Measurements of their underlying electronic structure, however, are still lacking. Using angle-resolved photo\-electron spectroscopy, we measure the valence band dispersion of single-crystal CH$_3$NH$_3$PbBr$_3$. The dispersion of the highest energy band is extracted applying a modified leading edge method, which accounts for the particular density of states of organic-inorganic perovskites. The surface Brillouin zone is consistent with bulk-terminated surfaces both in the low-temperature orthorhombic and the high-temperature cubic phase. In the low-temperature phase, we find a ring-shaped valence band maximum with a radius of 0.043~{\AA}$^{-1}$, centered around a 0.16~eV deep local minimum in the dispersion of the valence band at the high-symmetry point. Intense circular dichroism is observed. This dispersion is the result of strong spin-orbit coupling. Spin-orbit coupling is also present in the room-temperature phase. The coupling strength is one of the largest reported so far.

\end{abstract}

\maketitle

Organic-inorganic perovskite compounds (OIPCs) have potential applications in opto\-electronics ranging from high-efficiency thin film solar cells~\cite{Yang2015, saliba2016, stranks2015} to photo\-detectors~\cite{dou2014} and scintillators~\cite{Yakunin2015}, and from optical refrigeration~\cite{ha2015} to low-threshold nano\-lasers~\cite{zhu2015}. Yet, fundamental questions remain open concerning the electronic structure underlying their favorable photo\-transport properties. Relativistic effects, i.~e. spin-orbit coupling (SOC) and resulting spin splitting, are expected because of the constituting heavy elements~\cite{even2013, even2014, kim2014, brivio2014, umari2014}. Spin splitting could be strong enough to contribute to the long carrier lifetimes in OIPCs~\cite{zheng2015, etienne2016, azarhoosh2016}, and to allow for  OIPC-based spintronic devices~\cite{kepenekian2015, Li2016}. However, the spin splitting found in calculations~\cite{park2015, quarti2014, motta2015} is extremely sensitive to the orientation of the organic cation and to distortions of the inorganic cage, with calculated Rashba parameters of energetically similar structures~\cite{quarti2014} ranging from $<0.1$~eV~{\AA} to almost $10$~eV~{\AA}. As direct measurements of the electronic structure are lacking~\cite{berry2015}, the actual extend of Rashba splitting in OIPCs remains unknown. Yet, intense circular dichroism in pump-probe spectroscopy~\cite{giovanni2015} 
and spin dependence of charge dissociation and recombination at room temperature~\cite{hsiao2015} in CH$_3$NH$_3$PbI$_3$ hint to the possibility to create spin polarization in OIPCs.

\begin{figure}
\center\includegraphics[width=0.45\textwidth]{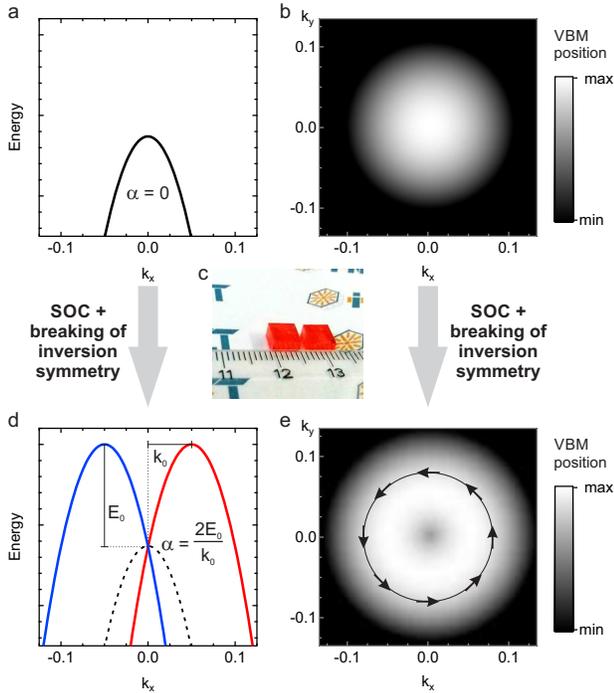}
\caption{Influence of spin-orbit coupling (SOC) and the resulting Rashba effect on VB dispersion. (a,~b)~Without SOC, a doubly spin-degenerate band with a single maximum is expected. (d,~e)~In the presence of SOC and inversion-symmetry breaking fields, it splits up into two spin-split bands. The strength of SOC is quantified by the Rashba parameter $\alpha$. Arrows in (e) depict the orientation of the in-plane component of the spin. (c)~shows the samples under investigation.
}	 
\label{fig1}
\end{figure}

We report measurements of the electronic structure of single-crystal CH$_3$NH$_3$PbBr$_3$ using angle-resolved photo\-electron spectroscopy (ARPES). CH$_3$NH$_3$PbBr$_3$ grows large cubic single crystals~\cite{shi2015, dong2015}, as shown in figure~\ref{fig1}, suitable for cleaving in ultrahigh vacuum and for ARPES experiments. 
Applying a leading edge method that takes into account the density of states (DOS) of the OIPC, we extract the dispersion of the highest-energy valence band (VB). We find a spin-splitting of the band, quantified by the Rashba parameter $\alpha$. The effect of Rashba splitting on band dispersion is illustrated in figure~\ref{fig1}. Rashba splitting arises when orbitals with SOC are subject to symmetry-breaking electric fields. It causes a splitting of a doubly spin-degenerate band into two bands shifted with respect to each other in k-space by $k_0$. At the central high-symmetry point, a minimum arises of depth $E_0$. For CH$_3$NH$_3$PbBr$_3$ we find  Rashba parameters $\alpha = 2E_0/k_0$ of $7\pm1$~eV~{\AA} and $11\pm4$~eV~{\AA} in the orthorhombic and the cubic phase, respectively. These values are amongst the highest ones reported~\cite{Ishizaka2011}. Our findings point out the critical role of 
local inversion-symmetry breaking fields 
in OIPCs, and of the resulting spin splitting~\cite{park2015, quarti2014, motta2015}. We hope our study will stimulate further investigations of spin structure in OIPCs to elucidate the role of spin splitting and possible spin polarization in existing and future applications.

Crystal were synthesized according to the modified procedure in reference~\onlinecite{shi2015}. For details of sample preparation see figure~S1 of the Supplemental Material~\cite{suppl}. 
Resulting CH$_3$NH$_3$PbBr$_3$ single crystals with sizes of $\approx0.5$~cm are shown in figure~\ref{fig1}~(c). 
They exhibit a cubic shape, exposing (001)-oriented facets. Clean surfaces were prepared by cleaving crystals in vacuum (base pressure $10^{-6}$~Pa) parallel to the exposed (001) face. This results in a single, optically flat terrace with an area of several~mm$^2$. Samples were then transferred to ultrahigh vacuum (base pressure $2\cdot10^{-8}$~Pa) within 1~min. We monitor possible loss of methylamine using thermal desorption spectroscopy. No desorption of the organic compound was detectable  
below 320~K, see figure~S2~(a)~\cite{suppl}. 
ARPES spectra were recorded using unpolarized light with a photon energy of 21.2~eV (He~I). Satellite lines were subtracted. The fourth harmonic of a femtosecond Ti:Sa~oscillator (h$\nu=6.2$~eV) was used for laser-ARPES. Spectra were recorded with an ellipsoidal display-type analyzer~\cite{Schnell1984}. It collects two-dimensional angular photoelectron distribution patterns from an acceptance cone covering $-35^\circ$ to $45^\circ$ both along and perpendicular to the plane of light incidence without the need to move sample or analyzer. 
During cooling, we perform in situ using photoluminescence spectroscopy 
reproducing the phase transitions reported in literature~\cite{swainson2015}, see figure~S2~(b)~\cite{suppl}.

\begin{figure}
\center\includegraphics[width=0.99\columnwidth]{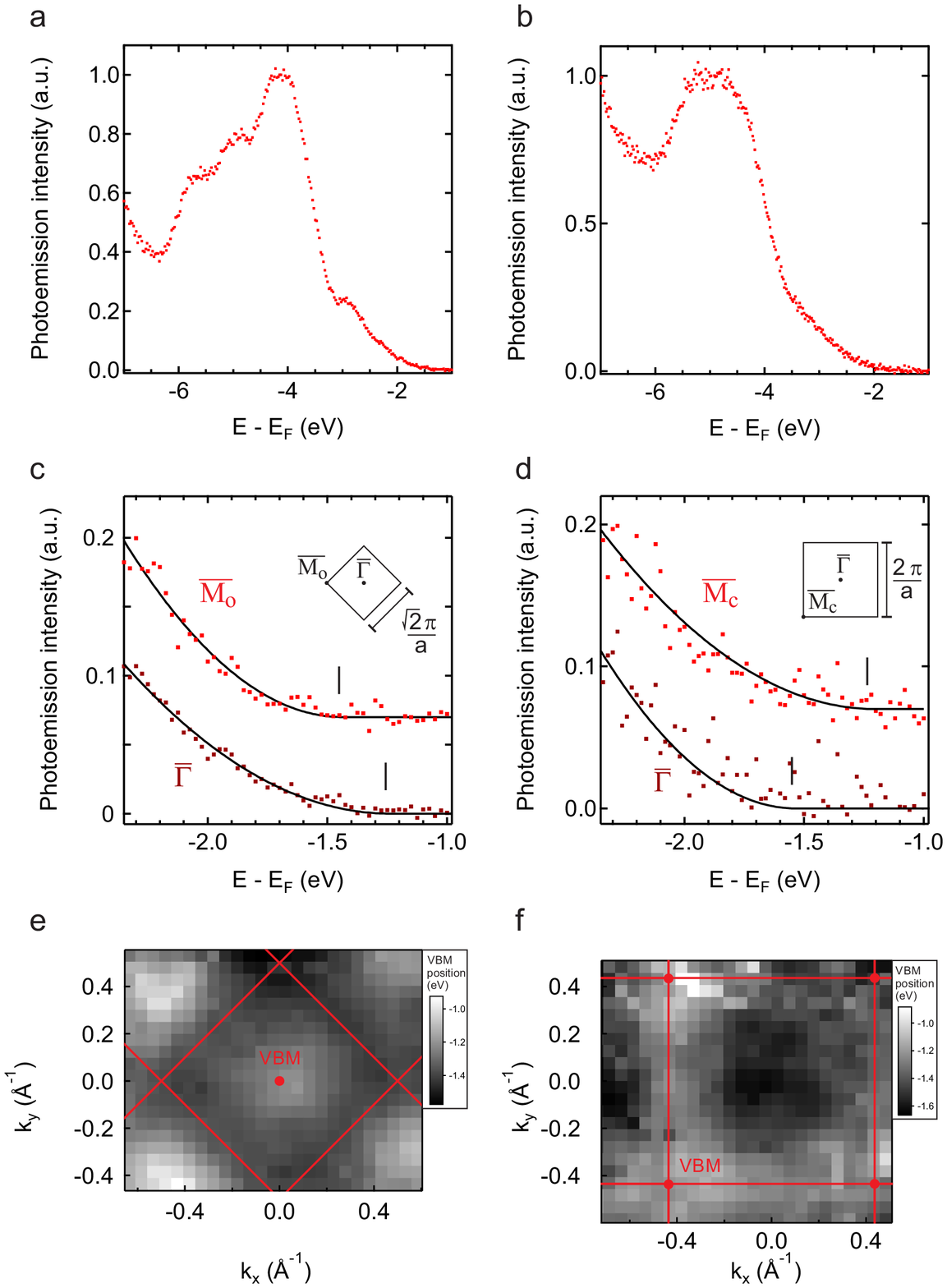}
\caption{ARPES data from single crystal CH$_3$NH$_3$PbBr$_3$ in the (a, c, e)~low-temperature temperature orthorhombic and the (b, d, f)~room-temperature cubic phase. (a, b) Normal emission ARPES spectra. (c, d)~Close-up of the VBM region of ARPES spectra recorded at the centers $\mathrm{\overline{\Gamma}}$ and the corners $\mathrm{\overline{M}_o}$ and $\mathrm{\overline{M}_c}$ of the orthorhombic and the cubic surface Brillouin zone, respectively. Bulk-terminated surface Brillouin zones for both phases are illustrated in the insets. The position of the VBM is determined by parabolic fits to the data (solid lines). Spectra from the Brillouin zone boundary are offset by 0.07 for reasons of clarity. (e, f)~Position of the highest-energy VB as a function of parallel momentum. 
}	 
\label{fig2}
\end{figure}

Figure~\ref{fig2}~(a) shows a normal emission photoelectron spectrum of CH$_3$NH$_3$PbBr$_3$ in the low-temperature orthorhombic phase. The VBs extend roughly from $-6$~eV to $-1$~eV, with a width of 5~eV matching the one reported from calculations~\cite{mosconi2013, park2015}. The doping level exhibits sample-to-sample variations from intrinsic, as shown in figure~\ref{fig2}~(a), to p-type with the VB maximum (VBM) close to the Fermi level, as in figure~\ref{fig4}. Doping acts as a rigid shift of the spectra, and otherwise does not affect the results reported here. 
The room-temperature normal emission spectrum of  CH$_3$NH$_3$PbBr$_3$ is shown in Figure~\ref{fig2}~(b). Substructures appear less pronounced than at low temperature. The widths of the spectra are similar. 

As known from calculations~\cite{mosconi2013, park2015, endres2016}, the highest-energy VB has a significantly smaller effective mass than deeper-lying bands, and contributes little to the DOS. The DOS is approximately parabolic on a scale  of  several hundred meV around the band edges~\cite{jishi2014, umari2014, motta2015, menendez2014}, in contrast to the typical square-root behavior of three-dimensional semiconductors with parabolic bands. Accounting for this particular DOS, we use a modified leading-edge method to determine the position of the highest-energy VB. The VB edge is approximated by a parabolic fit to the ARPES spectra, reflecting the DOS found in calculations. The procedure is illustrated in Figure~\ref{fig2}~(c) and~(d) for CH$_3$NH$_3$PbBr$_3$ in the low-temperature and room-temperature phase. 
The fitting curve is set to zero above the vertex, giving the position of the topmost VB. At the VBM, photo\-emission intensity drops to zero. We find no indication of in-gap surface electronic states. 
The low trap density allows a detailed analysis of the VB edge, which was not possible in a previous study on CH$_3$NH$_3$PbI$_3$ thin films~\cite{wu2015} with surface traps all the way up to the Fermi level. 

The insets in figure~\ref{fig2}~(c) and~(d) show the surface Brillouin zones expected for bulk-terminated (001)~surfaces in both phases. Here $a=6$~{\AA}~\cite{poglitsch1987, mashiyama1998} is the cubic lattice constant. The low-temperature Brillouin zone is approximated by a square, since 
domains of different orientation can form upon cooling the crystal, which are not resolved by our experiment (spatial resolution: 3~mm). 
The given surface Brillouin zone reflects domains with out-of-plane alignment of the ortho\-rhombic $b$~axis.

The position of the VB  edge for various parallel momenta is shown in figure~\ref{fig2}~(e) and~(f)  for both phases under investigation. The number of k-points was reduced for reasons of clarity. 
At low temperature, band maxima are located at normal emission, i.~e. at~$(0,0)$, and at around $(\pm0.5,\pm0.5)$~{\AA}$^{-1}$. Their periodicity is consistent with the bulk-terminated surface Brillouin zone shown in figure~\ref{fig2}~(c) and a reciprocal lattice constant of $\sqrt{2}\cdot\frac{\pi}{a}=0.7$~{\AA}$^{-1}$. No indications of surface reconstructions are found. Also, we  do not observe band maxima at the $\mathrm{\overline{M}_o}$ points, which would be the signature of domains with in-plane orientation of the $b$~orthorhombic axis. The observed out-of-plane alignment of the $b$~axis at the surface is consistent with scanning tunneling microscopy~\cite{ohmann2015}. 
In the room-temperature phase, figure~\ref{fig2}~(f), only the band maxima at the boundaries of the Brillouin zone remain, whereas its center forms a minimum. 
This is consistent with the bulk terminated surface of the cubic crystal as shown in the inset of figure~\ref{fig2}~(d), without indications of surface reconstructions.

\begin{figure}
\center\includegraphics[width=0.99\columnwidth]{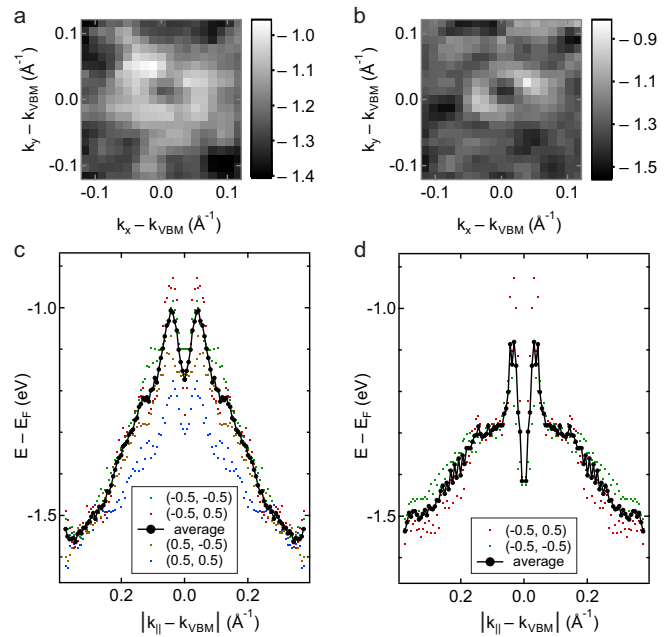}
\caption{Close up of the dispersion of the highest-energy VB in the surrounding of its maximum for the (a) low-temperature and (b) room-temperature phase. (c,~d) band dispersion after azimuthal averaging for both phases. Dots give the dispersion in the vicinity of individual high-symmetry points. Black symbols depict their average.
}	 
\label{fig3}
\end{figure}

\begin{figure*}
\center\includegraphics[width=0.89\textwidth]{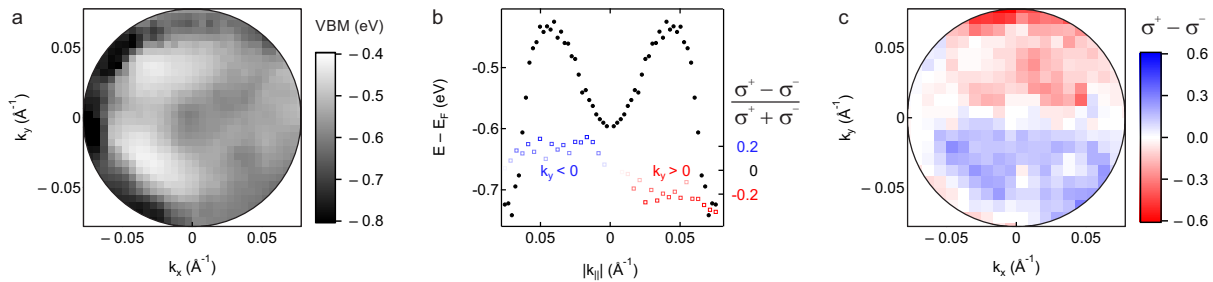}
\caption{(a) Dispersion of the central band maximum of the low-temperature orthorhombic phase from laser-ARPES (h$\nu=6.2$~eV). (b)~VB dispersion (black filled symbols, left axis) and circular dichroism signal (blue and red squares, right axis) after azimuthal averaging. Azimuthal averaging of the circular dichroism signal was performed separately for $k_y < 0$ and $k_y > 0$, or the lower and upper half of (c), respectively. (c) Circular dichroism  signal, showing the difference in the intensity distribution of photoelectrons after excitation with right- and left-handed circularly polarized light.
}	 
\label{fig4}
\end{figure*}

Figures~\ref{fig3}~(a) and (b) show close-ups of the band maxima. 
Data for equivalent maxima were averaged for better statistics. Figures~\ref{fig3}~(c) and~(d) show the photo\-emission intensity after averaging along the azimuthal direction around the high-symmetry points. Data were mirrored with respect to $\left| k_{||} -  k_{VBM}\right| = 0$ for reasons of clarity. Dots give the positions of individual band maxima,  black symbols depict their average. Since the experimental resolution is highest for the maxima at \mbox{(-0.5,~$\pm0.5$)}~{\AA}$^{-1}$, which are found closest to the central axis of the analyzer in our experiments, only these maxima are taken into account in the averaging and for the room-temperature phase.

Independent of structural phase, all maxima exhibit a sub-structure with a local minimum at the high symmetry points and a surrounding, approximately ring-shaped maximum. The band maximum is shifted from the high symmetry point by $k_0=0.043\pm0.002$~{\AA}$^{-1}$. Band maxima in the orthorhombic phase are found at an energy of $E_o=0.16\pm0.04$~eV with respect to the minimum at $k=0$, and at $E_c=0.24\pm0.08$~eV in the cubic phase, respectively. The spacing between the maxima is too large to result from orthorhombic domains of different orientation, or respective local tetragonal domains at room temperature, even if these inhomogeneities are present at the surface. We assign the shift of the band maxima in k-space to spin-orbit coupling in the material. 

To test this interpretation, we performed laser-ARPES using circularly polarized light. Circular dichroism is a necessary consequence of spin-orbit coupling. Because of the low photon energy of the laser (h$\nu=6.2$~eV), the technique is limited to the VBM at normal emission of the low-temperature phase. Since ARPES signal from the VB overlaps with secondary electrons at low kinetic energies, an exponential background was introduced in the fitting procedure in addition to the parabolic VB DOS. Figures~\ref{fig4}~(a,~b) show the resulting 
dispersion. Data analysis is restricted to $\left|k_{||}\right| < 0.08$~{\AA}$^{-1}$ because of the low photon energy. Again, a ring-shaped maximum with $k_0=0.043\pm0.005$~{\AA}$^{-1}$ and $E_o=0.16\pm0.02$~eV is found. The dispersion matches the one from the conventional ARPES experiment in figure~\ref{fig3}~(c). Notably, the information depth in laser-ARPES is a factor 15 larger because of reduced photon energy~\cite{seah1979} and emission angle. A circular dichroism map is given in figure~\ref{fig4}~(c). It shows the difference~$\sigma^+ - \sigma^-$ in photo\-emission intensity from the VB in experiments with right- and left-handed circularly polarized light. The anti\-symmetric structure expected for spin-orbit split states is found for $\left| k_{||} \right| < 0.05$~{\AA}$^{-1}$. The amplitude of the circular dichroism signal $\left|\frac{\sigma^+ - \sigma^-}{\sigma^+ + \sigma^-}\right|$ is $0.2\pm0.05$.

The most simple model accounting for SOC is the Rashba Hamiltonian~\cite{rashba1960, casella1960}. Its solutions $E^\pm = \hbar^2k^2/2m^*\pm\alpha \left|k\right|$ are two parabolas offset in k-space. The parameters are the effective mass $m^*$ of the bands and the Rashba parameter $\alpha$. The measured VB dispersion of CH$_3$NH$_3$PbBr$_3$ 
deviates from a parabola, but is closer to a hyperbola or even a cone, consistent with the parabolic DOS~\cite{jishi2014, umari2014, motta2015, menendez2014}. 
The resolution of our experiment is insufficient to reliably extract effective masses at the maximum of the bands. The parameter $\alpha = 2 E_0/k_0$, however, can be determined consistently
. We find $\alpha_o=7\pm1$~eV~{\AA} and $\alpha_c=11\pm4$~eV~{\AA} for the orthorhombic and the cubic phase. These range at the higher end of calculated values for OIPCs~\cite{quarti2014} which give $\alpha=0.1$...$10$~eV~{\AA}. Since ARPES is a surface-sensitive technique, the reduced symmetry at the surface may enhance non-centrosymmetric fields and the measured Rashba splitting. In the bulk, CH$_3$NH$_3$PbBr$_3$ forms an ordered orthorhombic structure of $Pnma$ symmetry at  low temperature~\cite{Chi2005, mashiyama2007},  and a disordered cubic structure with an average $Pm3m$ symmetry at room temperature~\cite{mashiyama1998, Swainson2003, swainson2015}. Both structures are centro\-symmetric after averaging over a length scale probed by x-ray and neutron diffraction, with no Rashba-type spin-splitting expected. Scanning tunneling microscopy (STM) on (001) surfaces of CH$_3$NH$_3$PbBr$_3$ reveals that at the surface both centro\-symmetric (anti\-ferroelectric) and non-centro\-symmetric (ferro\-electric) domains form after cooling the crystal from room temperature to low temperature~\cite{ohmann2015}. Calculations showed that the non-centro\-symmetric structure has the lower band gap. We thus speculate that it might dominate our ARPES data, which are spatially averaged over a spot size of (3~mm)$^2$. It remains an open question how deeply the surface domains extend into the bulk. Freezing of local disorder was also found in photo\-luminescence~\cite{galkowski2016a, phuong2016, panzer2016} and transient absorption spectroscopy~\cite{phuong2016} of thin films of CH$_3$NH$_3$PbI$_3$, which has been ascribed to coexisting orthorhombic and tetragonal domains. We hypothesize that for thin films, the observed crystalline alignment at the surface may have a significant impact on the structure throughout the film at low temperature. Less is known about the surface structure at room temperature. Piezo\-force response in atomic force microscopy at room temperature was detected at the surface  of tetragonal CH$_3$NH$_3$PbI$_3$ and ascribed to static ferro\-electric ordering~\cite{coll2015, kutes2014}, but not in cubic CH$_3$NH$_3$PbBr$_3$~\cite{zhao2015}, implying disorder at the latter surface. The underlying bulk lattice is cubic and centro\-symmetric in average. Yet, the individual bromide atoms reside 0.04~$a$ from the high-symmetry sites in a disordered arrangement~\cite{mashiyama1998, beecher2016, yaffe2016}. In CH$_3$NH$_3$PbI$_3$, the methylamine ion wobbles and flips on 0.3~ps and 3~ps time scales~\cite{bakulin2015}, and similar motions may be expected for CH$_3$NH$_3$PbBr$_3$. The resulting spatially  and temporally fluctuating fields give rise to a dynamical Rashba effect. Calculations report an average $\alpha=4$~eV~{\AA} for bulk cubic CH$_3$NH$_3$PbI$_3$~\cite{etienne2016}. Notably, Rashba splitting results in an energy upshift of the band maximum by $E_0$ with respect to the case of $\alpha=0$, as illustrated in figure~\ref{fig1}~(d). The way our data are analyzed, we measure the dispersion of the highest-lying bands, giving an upper limit for spin-splitting in the disordered system.

To the best of our knowledge, the Rashba splitting reported here is the strongest one found experimentally for any system. For layered BiTeI, $\alpha=3.8$~eV~{\AA} has been reported~\cite{Ishizaka2011}. The exceptionally large $\alpha$ we extract may in part be the result of deviations in dispersion from the parabolic one of the Rashba model.  Nevertheless, with $\hbar k_0$ being larger than the momentum of visible photons and $E_0 > kT$, spin-splitting puts constraints to optically allowed transitions and carrier recombination in OIPCs, thus forming a crucial ingredient to photo\-transport in the material. Enhancement of carrier lifetimes by one to two orders of magnitude were predicted~\cite{zheng2015, azarhoosh2016} for CH$_3$NH$_3$PbI$_3$ and a static Rashba splitting in the conduction band of $E_0=0.1$...$0.2$~eV. The VB Rashba splitting is mediated by the SOC of the Pb~$6p$ orbital~\cite{kim2014} for both CH$_3$NH$_3$PbI$_3$ and CH$_3$NH$_3$PbBr$_3$, but is sensitive to details of the crystal structure~\cite{amat2014, quarti2014}. In most OIPCs, the Rashba splitting is stronger in the conduction band than in the VB~\cite{umari2014, amat2014, etienne2016}. 
Calculations also report a complex spin structure involving both Rashba and Dresselhaus spin splittings~\cite{kepenekian2015, pedesseau2016}. We hope our results will stimulate further studies on the effect of spin splitting and polarization on the electronic properties of this multifaceted system.

In summary, we report measurements of the electronic structure of single crystal CH$_3$NH$_3$PbBr$_3$ in the orthorhombic and the  cubic phase. Using a modified leading edge method, which accounts for the DOS of the OIPC, we extract the dispersion of the highest-energy VB. Surface Brillouin zones are consistent with bulk terminated surfaces. In the low-temperature phase, the orthorhombic $b$~axis shows preferential out-of-plane alignment at the surface. No indication of surface electronic states within an energy range of 1~eV above the VB maximum is found. 
Most importantly, we find a splitting of the VBM in k-space by $k_0=0.043$~{\AA}$^{-1}$, giving rise to a minimum around 0.16~eV deep at the high symmetry points. Together with the observed circular dichroism, the dispersion indicates a Rashba splitting that is amongst the strongest reported.

\bibliography{perovskites}

\end{document}